\documentclass{article}
\usepackage{amsfonts}
\usepackage{mathrsfs}
\usepackage{amsmath}
\def\R{\mathbb{R}}

\def\<{\langle}

\def\R{{\mathbb{R}}}

\def\C{{\mathbb{C}}}\def\K{{\mathcal{K}}}
\def\>{\rangle}
\def\EQB{\begin{equation}}
\def\EQN{\end{equation}}
\numberwithin{equation}{section}

\date{ }
\begin{document}
\begin{center}
\begin{minipage}{150mm}
%\vskip 0.2in
{\bf\LARGE Time evolution and adiabatic approximation\\ in $PT$-symmetric quantum mechanics}\\

{Zhihua Guo and Huaixin Cao\\
\it\small College of Mathematics and Information Science,\\
\it\small Shaanxi Normal University, Xi'an 710062, China\\
\it\small Email: zhguosx@gmail.com, caohx@snnu.edu.cn} \vskip 0.1in

In this paper, we discuss time evolution and adiabatic approximation
in $PT$-symmetric quantum mechanics. we give the time evolving
equation for a class of $PT$-symmetric Hamiltonians and some
conditions of the adiabatic approximation for the class of
$PT$-symmetric Hamiltonians.

\vskip 0.1in
DOI:\  \hfill PACS numbers: 03.67.Mn, 03.65.Ud

\end{minipage}
\end{center}

\section{Introduction}
PT-symmetry theory [1] was proposed by Bender and collaborators in 1998, where they considered certain classes of Hamiltonians which seem not Hermitian in Hilbert spaces but having real spectra. Now, this theory has been widely discussed and developed [2-17]. It is well-known that in a conventual quantum mechanics, the time evolution of the system is described by the Schr$\ddot{o}$dinger equation of a Hamiltonian, which is a densely-defined Hermitian operator in Hilbert spaces. The Hermiticity of this Hamiltonian ensures that its spectra are real and time evolution is unitary. It is remarkable that the Hermiticity of a Hamiltonian  is not necessary for the reality of spectra. Typical examples are Hamiltonians of the form $H=p^2+x^2(ix)^\epsilon$, where $\epsilon\in \R$. When $\epsilon\geq0$, the spectrum of $H$ is real and positive as a consequence of $PT$-symmetry while when $\epsilon\in (-1,0)$, the eigenvalues are coming in complex conjugate pairs because the $PT$-symmetry is broken. Recently, a mathematical groundwork on this theory in [18]. Especially, for a $PT$-symmetric Hamiltonian on a Hilbert space, concepts of $PT$-frames, $CPT$-frames are introduced and  discussed.

In addition, as one of the oldest theorem in conventual quantum mechanics, the adiabatic theorem [19] tells us that consider a state evolving according to the Schr$\ddot{o}$dinger equation described by a Hamiltonian $H(t)(t\in [0,T]$, $T$ being the total evolving time) with eigenstates $\{|\psi_n(t)\rangle\}$ and corresponding eigenvalues $\{E_n(t)\}$, if the initial state is the $k$th-eigenstate $|\psi_k(0)\rangle$ and $H(t)$ varies slowly enough, the instantaneous state $|\psi(t)\rangle(t\in [0,T])$ of the system will remain close to the state $|\psi_k(T)\rangle$ at the end of the process. Based on the adiabatic approximation, this theorem has far-ranging application in many areas such as Landau-Zener transition in molecular physics [20], quantum field theory [21], geometric phase [22], geometric quantum computation [23] and new quantum algorithm [24]. The ``enough slow evolution" leads to a lot of  scholars' interests.  Tong et al. in [25-27] discussed the sufficiency and necessity of the quantitative condition for the validity of the adiabatic approximation. In 2008, A. Ambainis and O. Regev in [28] gave an elementary proof of the quantum adiabatic theorem. In 2011, J. E. Avron et al. in [29] established adiabatic theorems for generators of contracting evolutions.
Very recently, we introduced in [30] a function  in terms of eigenvalues and eigenstates of a time-dependent Hamiltonian on an arbitrary dimensional Hilbert space in conventual quantum mechanics and described quantitatively the slow evolution of the system.

In this paper, we will discuss time evolution and adiabatic
approximation in $PT$-symmetric quantum mechanics. In sec. 2, we
introduce the $CPT$-Frames in a Hilbert space. Then we will discuss
the evolution equation in $PT$-symmetric quantum mechanics in sec. 3
and adiabatic approximation in sec. 4. At last, we consider an
example as an application.

\section{$CPT$-Frames}
\setcounter{equation}{0}

Let $P$ be a bounded linear operator on a complex Hilbert space $\K$ and $T$ a bounded anti-linear (conjugate linear) operator on $\K$. If
the conditions $P^2 = T^2 = I$ and $PT = TP$ are satisfied,  then we call the pair $\{P,T\}$ a  \emph{$PT$-frame} on $(\K,\<\cdot|\cdot\>)$([30]). Let $\{P,T\}$ be the given $PT$-frame on $\K$. A linear operator $H$ in $\K$ is said to be \emph{$PT$-symmetric} ([30]) if it commutes with $PT$, i.e.,
$$[H,PT]= HPT-PTH = 0.$$ If $C$ is a bounded linear operator on $\K$, then the triple $\{C,P,T\}$ is said to be a \emph{$CPT$-frame} on $\K$ if the following conditions are satisfied.
(1) $CPT=TPC, C^2=I$, and
(2) $PC$ is positive definite with respective to the inner product $\langle\cdot|\cdot\rangle$ on $\K$, i.e.,
$\langle PCx|x\rangle\geq 0$ for all $x\in \K$; and $\langle PCx|x\rangle=0\Leftrightarrow x=0.$
A $CPT$-frame $\{C,P,T\}$ on $\K$ is said to be a \emph{$CPT$-frame} ([30]) for an operator $H$ in $\K$ if $CH=HC$.

Let $\{C,P,T\}$ be a $CPT$-frame on $\K$. Then we can obtain a positive definite inner product $(\cdot|\cdot)_{CPT}$ on $\K$ by
$$(x|y)_{CPT}=\<x|PCy\>,\ \ \forall x,y\in \K \eqno{(1)}$$
called a $CPT$-inner product. Since $PC$ is invertible, $\K$ is also a Hilbert space with respect to the $CPT$-inner product. For a densely defined linear operator $A$ in $\K$, we use the notation $A^{CPT}$ to denote the adjoint of $A$ with respect to the $CPT$-inner product. Clearly, $A^{CPT}=(PC)^{-1}A^\dag (PC)$. In the case that  $A^{CPT}=A$, we say that $A$ is $CPT$-Hermitian. Hence, $A$ is $CPT$-Hermitian if and only if $A^\dag PC=PCA$ if and only if
$\<Ax|PCy\>=\<x|PCAy\>$ for all $x,y\in D(A).$ For example, $PC$ is $CPT$-Hermitian; $C$ is $CPT$-Hermitian if and only if $P^\dag=P$. Moreover, a $CPT$-frame  $\{C,P,T\}$  on $\K$ is a $CPT$-frame for the operator $H_{a,b}:=aI+bC (a,b\in \R)$, and  $H_{a,b}$ is $PT$-symmetric. A linear operator $H$ in $\K$ is said to have {\it unbroken $PT$-symmetry} if it is $PT$-symmetric and every eigenvector (i.e., eigenstate) of $H$ is an eigenvector of $PT$. It is easy to check that eigenvalues of an operator that has unbroken $PT$-symmetry are all real([30]).

\section{Evolution equation in $PT$-symmetric quantum mechanics}
In this section, we will consider time evolution in $PT$-symmetric quantum mechanics.  To do this, we assume that for every $t\ge0$, $\{C(t),P,T\}$ is a $CPT$-frame  on $\K$, $H(t)$ is a linear operator in $\K$ which has unbroken $PT$-symmetry and is $C(t)PT$-Hermitian. Thus, we get a family of positive definite inner products $(\cdot|\cdot)_{C(t)PT}(t\ge 0)$ on $\K$. The norms $\|\cdot\|_{C(t)PT}$ induced by these inner products are time dependent and then dynamical, but they are all equivalent to the original norm $\|\cdot\|$ on $\K$ since
$$\|C(t)P\|^{-1/2}\cdot\|x\|\le\|x\|_{C(t)PT}\le\|PC(t)\|^{1/2}\cdot\|x\|,\ \ \forall t\ge0, \forall x\in\K. \eqno(2)$$

Next we want to discuss the evolution described by the following equation: $$i\hbar \frac{d}{dt}\phi(t)=H(t)\phi(t) (t\ge0).\eqno(3)$$

According to conventual quantum mechanics, the time evolution should be unitary, i.e., $\frac{d}{dt}(\phi(t)|\phi(t))_{C(t)PT}=0$ for all $t\ge0$ whenever $\phi(t)$ is a solution to (3). Denote $\frac{d}{dt}\phi(t)=\dot{\phi}(t)$ for convenience. Let $\phi(t)$ be a solution to (3). Then
\begin{eqnarray*}\frac{d}{dt}(\phi(t)|\phi(t))_{C(t)PT}&=&\frac{d}{dt}\<\phi(t)|PC(t)\phi(t)\>\\
&=&\<\dot{\phi}(t)|PC(t)\phi(t)\>+\<\phi(t)|P\dot{C}(t)\phi(t)\>+\<\phi(t)|PC(t)\dot{\phi}(t)\>\\
&=&-\frac{1}{i\hbar}\<H(t)\phi(t)|PC(t)\phi(t)\>+\<\phi(t)|P\dot{C}(t)\phi(t)\>+\frac{1}{i\hbar}\<\phi(t)|PC(t)H(t){\phi}(t)\>\\
&=&-\frac{1}{i\hbar}(H(t)\phi(t)|\phi(t))_{C(t)PT}+\<\phi(t)|P\dot{C}(t)\phi(t)\>+\frac{1}{i\hbar}(\phi(t)|H(t){\phi}(t))_{C(t)PT}\\
&=&\<\phi(t)|P\dot{C}(t)\phi(t)\>.\end{eqnarray*}
Let $S_H$ be the set of all solution to Eqn.(3). Then we obtain the following conclusion which gives a characterization of unitary evolution of Eqn.(3).

{\bf Theorem 1.} {\it The time evolution of Eqn.(3) is unitary if and only if $\<\phi(t)|P\dot{C}(t)\phi(t)\>=0$ for all $\phi\in S_H$ and all $t\ge0$.}

From this theorem, we get the following corollary.

{\bf Corollary 1.} {\it If for all $t\ge0$, $\dot{C}(t)=0$, then the time evolution of Eqn.(3) is unitary.}

{\bf Example 1} We suppose that $\{C,P,T\}$ is a $CPT$-frame  on $\K$, $a(t)$ and $b(t)$ are any real-valued functions on the interval $[0,\infty)$. Define  $C(t)=C, H(t)=a(t)I+b(t)C$ for all $t\ge0$. Then $\{C(t),P,T\}$ is a $CPT$-frame for $H(t)$ for all $t\ge0$. Since $\dot{C}(t)=0$ for all $t\ge0$, Corollary 1 yields that
the time evolution of Eqn.(3) given by this Hamiltonian is unitary.

 Generally, $\<\phi(t)|P\dot{C}(t)\phi(t)\>$ is not necessarily identically equal to zero on the interval $[0,\infty)$.
Thus, we consider the evolution described by the following equation:
$$i\hbar \dot{\phi}(t)=\left(H(t)+iG(t)\right)\phi(t) (t\ge0),\eqno(4)$$
where $G(t)$ is a $C(t)PT$-Hermitian operator in $\K$ for all $t\ge0$.  Let $\phi(t)$ be a solution to (4). Then
\begin{eqnarray*}\frac{d}{dt}(\phi(t)|\phi(t))_{C(t)PT}&=&\frac{d}{dt}\<\phi(t)|PC(t)\phi(t)\>\\
&=&\<\dot{\phi}(t)|PC(t)\phi(t)\>+\<\phi(t)|P\dot{C}(t)\phi(t)\>+\<\phi(t)|PC(t)\dot{\phi}(t)\>\\
&=&-\frac{1}{i\hbar}\<H(t)\phi(t)|PC(t)\phi(t)\>+\<\phi(t)|P\dot{C}(t)\phi(t)\>+\frac{1}{i\hbar}\<\phi(t)|PC(t)H(t){\phi}(t)\>\\
&&+\frac{1}{\hbar}\<G(t)\phi(t)|PC(t)\phi(t)\>+\frac{1}{\hbar}\<\phi(t)|PC(t)G(t){\phi}(t)\>\\
&=&\<\phi(t)|\Big(P\dot{C}(t)+\frac{2}{\hbar}PC(t)G(t)\Big)|{\phi}(t))\>.
\end{eqnarray*}

Let $S_{HG}$ be the set of all solution to Eqn.(4). Then we obtain the following conclusion which gives a characterization of unitary evolution of Eqn.(4).

{\bf Theorem 2.} {\it The time evolution of Eqn.(4) is unitary if and only if $\<\phi(t)|(P\dot{C}(t)+\frac{2}{\hbar}PC(t)G(t))\phi(t)\>=0$ for all $\phi\in S_{HG}$ and all $t\ge0$.}

From this theorem, we get the following corollary.

{\bf Corollary 2.} {\it The time evolution of the following equation is unitary.}
$$i\hbar \dot{\phi}(t)=\left(H(t)-\frac{i\hbar}{2}C(t)\dot{C}(t)\right)\phi(t) (t\ge0).\eqno(5)$$

Let $\phi(t)$ be a solution to Eqn.(5). Then $\frac{d}{dt}(\phi(t)|\phi(t))_{C(t)PT}=0$ for all $t\ge0$.
Consequently,
$$\|\phi(t)\|_{C(t)PT}=[(\phi(t)|\phi(t))_{C(t)PT}]^{1/2}=[(\phi(0)|\phi(0))_{C(0)PT}]^{1/2}=\|\phi(0)\|_{C(0)PT}$$
and then there exists a unitary operator $U(t):(\K,(\cdot|\cdot)_{C(0)PT})\rightarrow (\K,(\cdot|\cdot)_{C(t)PT})$ such that
$U(t)\phi(0)=\phi(t)$ for all $t\ge0$ with $U(0)=I$.  This shows that if Eqn.(5) is solvable, then for all $t\ge0$, Hilbert spaces $(\K,(\cdot|\cdot)_{C(0)PT})$ and $(\K,(\cdot|\cdot)_{C(t)PT})$ are unitarily isomorphic. Therefore, for all $s,t\ge0$, Hilbert spaces $(\K,(\cdot|\cdot)_{C(s)PT})$ and $(\K,(\cdot|\cdot)_{C(t)PT})$ are unitarily isomorphic. Furthermore, we obtain that
$$i\hbar \dot{U}(t)\phi(0)=\Big(H(t)-\frac{i\hbar}{2}C(t)\dot{C}(t)\Big)U(t)\phi(0).$$
This implies that if for every initial state $x_0\in\K$, Eqn.(5) has always solution $\phi(t)$ with $\phi(0)=x_0$, then
there exists a family of unitary operators $U(t):(\K,(\cdot|\cdot)_{C(0)PT})\rightarrow (\K,(\cdot|\cdot)_{C(t)PT})$ with $U(0)=I$ such that
$$i\hbar \dot{U}(t)=\Big(H(t)-\frac{i\hbar}{2}C(t)\dot{C}(t)\Big)U(t)(t\ge0).\eqno{(6)}$$

Conversely, if Eqn.(6) has a unitary solution $U(t):(\K,(\cdot|\cdot)_{C(0)PT})\rightarrow (\K,(\cdot|\cdot)_{C(t)PT})$ with $U(0)=I$, then for every initial state $x_0\in\K$, the function $\phi(t)=U(t)x_0$ is a solution to Eqn.(5) with $\phi(0)=x_0$.

As a conclusion, we obtain the following.

{\bf Theorem 3.} {\it For every initial state $x_0\in\K$ there exists a solution $\phi(t)$ with $\phi(0)=x_0$ to Eqn.(5) if and only if there exists a unitary solution $U(t)$ with $U(0)=I$ to Eqn.(6); in that case $\phi(t)=U(t)\phi(0)$.}

\section{$PT$-Symmetric adiabatic approximation}

In this section, we will discuss adiabatic approximation problem in $PT$-symmetric quantum mechanics.
In what follows, we assume that for every $t\ge0$, $\{C(t),P,T\}$ is a $CPT$-frame  on $\K$, $H(t)$ is a linear operator in $\K$ which has unbroken $PT$-symmetry and is $C(t)PT$-Hermitian.
For convenience, we denote $(\cdot,\cdot)_t=(\cdot,\cdot)_{C(t)PT}$ and $\|\cdot\|_t=\|\cdot\|_{C(t)PT}$.
Next, let us postulate that $H(t)$ has eigenstates $\psi_n(t)(n\in\Lambda )$ for eigenvalues $E_n(t)(n\in\Lambda )$, which consist an orthonormal  basis for $(\K,(\cdot,\cdot)_t)$. Thus,  $(\psi_m(t)|\psi_n(t))_t=\delta_{mn}$ for all $t\ge0$ and all $m,n\in\Lambda $. Since $H(t)$ has unbroken $PT$-symmetry, all $E_n(t)$ are real numbers.

Let $m\in\Lambda $ and $\theta(t)$ be a real-valued function on the interval $[0,\infty)$, and $\psi(t)=e^{i\theta(t)}\psi_m(t)$. Suppose that the function $\psi(t)$ is a solution to (5).  First, by an easy computation, we have
$$i\hbar\dot{\psi}(t)=i\hbar e^{i\theta(t)}\Big(i\dot{\theta}(t)\psi_m(t)+\dot{\psi}_m(t)\Big)$$
and
$$\Big(H-\frac{i\hbar}{2}C(t)\dot{C}(t)\Big)\psi(t)=e^{i\theta(t)}\Big(E_m(t)\psi_m(t)-
\frac{i\hbar}{2}C(t)\dot{C}(t)\psi_m(t)\Big).$$
Since $\psi(t)$ is a solution to (5), the right sides of the two equalities above are equal,
$$i\hbar \Big(i\dot{\theta}(t)\psi_m(t)+\dot{\psi}_m(t)\Big)=E_m(t)\psi_m(t)-
\frac{i\hbar}{2}C(t)\dot{C}(t)\psi_m(t).$$
Use the $C(t)PT$-inner product with $\psi_n(t)$ from the left-hand side, then
$$i\hbar \Big(i\dot{\theta}(t)\delta_{nm}+\langle\psi_n(t)|PC(t)\dot{\psi}_m(t)\rangle\Big)=E_m(t)\delta_{nm}-
\frac{i\hbar}{2}\langle\psi_n(t)|P\dot{C}(t)\psi_m(t)\rangle.$$
It implies that
$$\langle\psi_n(t)|PC(t)\dot{\psi}_m(t)\rangle=-\frac{1}{2}\langle\psi_n(t)|P\dot{C}(t)\psi_m(t)\rangle(n\neq m),\eqno(7)$$
 and
$$i\hbar \Big(i\dot{\theta}(t)+\langle\psi_m(t)|PC(t)\dot{\psi}_m(t)\rangle\Big)=E_m(t)-
\frac{i\hbar}{2}\langle\psi_m(t)|P\dot{C}(t)\psi_m(t)\rangle.\eqno(8)$$
Now applying the fact that
$$\langle\psi_m(t)|P\dot{C}(t)\psi_m(t)\rangle=-2\textmd{Re}\langle\psi_m(t)|PC(t)\dot{\psi}_m(t)\rangle$$
to Eqn.(8), then
$$\textmd{Im}\langle\psi_m(t)|PC(t)\dot{\psi}_m(t)\rangle=-\dot{\theta}(t)-\frac{1}{\hbar}E_m(t).$$
Equivalently,
$$\theta(t)=-\int_0^t\left(\frac{1}{\hbar}E_m(s)+\textmd{Im}\langle\psi_m(t)|PC(s)\dot{\psi}_m(s)\rangle\right)ds.\eqno(9)$$

Conversely, one can check that if Eqns.(7) and (9) hold, then $\psi(t)$ is a solution to Eqn.(5). In a word, we have the following.

{\bf Theorem 4.} {\it Let $m\in\Lambda $ and $\theta(t)$ be as in (9). Then the function $\psi(t)=e^{i\theta(t)}\psi_m(t)$ is a solution to (5) if and only if Eqn.(7) holds.}

{\bf Remark} When $C(t)\equiv C$ is independent of $t$, we have $\textmd{Re}\langle\psi_n(t)|PC\dot{\psi}_n(t)\rangle=-\frac{1}{2}\langle\psi_n(t)|P\dot{C}(t)\psi_n(t)\rangle=0$ and so  $$\textmd{Im}\langle\psi_n(t)|PC|\dot{\psi}_n(s)\rangle=-i \langle\psi_n(t)|PC\dot{\psi}_n(s)\rangle.$$
Set $\tilde{\psi}_n(t)=e^{i\int_0^t\left(-i\langle\psi_n(t)|PC|\dot{\psi}_n(s)\rangle\right)ds}\psi_n(t)$ for all $n\in\Lambda $.  Then $\tilde{\psi}_n(t)(\forall n )$ are eigenstates of $H(t)$ for eigenvalues $E_n(t)$ and consist an orthonormal basis for $\K$ satisfying $\langle\tilde{\psi}_n(t)|PC\dot{\tilde{\psi}}_n(t)\>=0$ for all $n$ and all $t\ge0$.
By using Theorem 4 for $\tilde{\psi}_n(t)$ in this case $\theta(t)=-\int_0^t\frac{1}{\hbar}E_m(s)ds$,  we know that
$$\psi(t)=e^{i\theta(t)}\tilde{\psi}_m(t)= e^{-i\int_0^t\left(\frac{1}{\hbar}+i\<\psi_m(s)|PC\dot{\psi}_m(s)\>\right)ds}\psi_m(t)$$ is a solution to (5) if and only if $\dot{\tilde{\psi}}_m(t)=$ for all $t\ge0$ if and only if $\langle\psi_m(t)|PC\dot{\psi}_m(t)\rangle\psi_m(t)+\dot{\psi}_m(t)=0$ for all $t\ge0$.

In the following, we consider the case that whether the operator-rotation $e^{iA(t)}\psi_m(t)$ is a solution to (5) whenever $\psi_m(t)$ is a solution to $i\hbar\dot{\psi}(t)=H(t)\psi(t)$.

{\bf Theorem 5} {\it Suppose that $\dot{C}(t)$ exists for all $t\ge0$, $\dot{C}(t)$, $E_m(t)$ and $H(t)$ are continuous on $[0,\infty)$. Put
$$ {A}(t)=\int_0^t\left(\frac{1}{\hbar}\Big(H(s)-E_m(s)I\Big)+\frac{i}{2}C(s)\dot{C}(s)\right)ds (\forall t\ge0),\eqno(10)$$
and $\psi(t)=e^{iA(t)}\psi_m(t)$. If $[A(t), H(t)]=0$ for all $t\ge0$, then the function $\psi(t)$ is a solution to (5)  if and only if      the eigenstate $\psi_m(t)$ of $H(t)$ is a solution to $i\hbar\dot{\psi}(t)=H(t)\psi(t)$.}

{\bf Proof. } From Eqn.(10) we have
$$H(t)=\hbar e^{-iA(t)}\dot{A}(t)e^{iA(t)}+E_m(t)I-\frac{i\hbar}{2}e^{-iA(t)}C(t)\dot{C}(t)e^{iA(t)}.$$
So
$$H(t)\psi_m(t)=\Big(\hbar e^{-iA(t)}\dot{A}(t)e^{iA(t)}+E_m(t)I-\frac{i\hbar}{2}e^{-iA(t)}C(t)\dot{C}(t)e^{iA(t)}\Big)\psi_m(t).\eqno(11)$$
By an easy computation, we can obtain from Eqn.(11) that
$$-\hbar\dot{A}(t)e^{iA(t)}\psi_m(t)+H(t)e^{iA(t)}\psi_m(t)
=E_m(t)e^{iA(t)}\psi_m(t)-\frac{i\hbar}{2}C(t)\dot{C}(t)e^{iA(t)}\psi_m(t).$$
On the other hand,
$$i\hbar\dot{\psi}(t)=-\hbar\dot{A}(t)e^{iA(t)}\psi_m(t)+e^{iA(t)}i\hbar\dot{\psi}_m(t),$$
\begin{eqnarray*}
\Big(H(t)-\frac{i\hbar}{2}C(t)\dot{C}(t)\Big)\psi(t)&=&E_m(t)e^{iA(t)}\psi_m(t)-\frac{i\hbar}{2}C(t)\dot{C}(t)e^{iA(t)}\psi_m(t)\\
&=&-\hbar\dot{A}(t)e^{iA(t)}\psi_m(t)+H(t)e^{iA(t)}\psi_m(t).
\end{eqnarray*}
This implies that $i\hbar\dot{\psi}(t)=\Big(H(t)-\frac{i\hbar}{2}C(t)\dot{C}(t)\Big)\psi(t)$ if and only if $ H(t)e^{iA(t)}\psi_m(t)=e^{iA(t)}i\hbar\dot{\psi}_m(t)$ if and only if $ H(t)\psi_m(t)=i\hbar\dot{\psi}_m(t)$. $\Box$

{\bf Remark 2} The condition that $[A(t), H(t)]=0$ for all $t\ge0$ seems to be strong, but it may be satisfied. For example,  the Hamiltonian $H(t)$ in Example 1 satisfies this condition. Moreover, we see from Theorem 5 that the eigenstate $\psi_m(t)$ solves approximately Eqn.(3) if and only if the function $\psi(t)=e^{iA(t)}\psi_m(t)$ solves approximately (5).

{\bf Theorem 6} {\it Let $\psi(t)$ be a solution to (5) with
$\psi(0)=\psi_m(0)$ and $0<\varepsilon<1$. Then $1-|(\psi_m(t)|\psi(t))_t|<\varepsilon$ for all $[0,T]$ provided that}
$$V(T):=\int_0^T\|(PC(s))^{1/2}\|\left(\|\dot{\psi}_m(s)\|+\frac{1}{2}\|C(s)\dot{C}(s)\psi_m(s)\|\right)ds<\varepsilon.\eqno(12)$$

{\bf Proof.} Let $\psi(t)=\sum_na_n(t)e^{i\theta_n(t)}\psi_n(t)$, where $\theta_n(t)=-\frac{1}{\hbar}\int_0^tE_n(s)ds$ and $\|\psi(t)\|_t=\left(\sum_n|a_n(t)|^2\right)^{1/2}=1$.
One can calculate directly that
$$i\hbar\dot{\psi}(t)
=i\hbar\sum_n\Big(\dot{a}_n(t)e^{i\theta_n(t)}\psi_n(t)-\frac{i}{\hbar}E_n(t)a_n(t)e^{i\theta_n(t)}\psi_n(t)
+a_n(t)e^{i\theta_n(t)}\dot{\psi}_n(t)\Big),\eqno(13)$$
$$\Big(H(t)-\frac{i\hbar}{2}C(t)\dot{C}(t)\Big)\psi(t)=\sum_n\Big(a_n(t)E_n(t)e^{i\theta_n(t)}\psi_n(t)
-\frac{i\hbar}{2}C(t)\dot{C}(t)a_n(t)e^{i\theta_n(t)}\psi_n(t)\Big).\eqno(14)$$
Since $\psi(t)$ is a solution to (5), combining Eqns.(13) and (14) gives
$$\sum_n\Big(\dot{a}_n(t)e^{i\theta_n(t)}\psi_n(t)+a_n(t)e^{i\theta_n(t)}\dot{\psi}_n(t)+
\frac{1}{2}C(t)\dot{C}(t)a_n(t)e^{i\theta_n(t)}\psi_n(t)\Big)=0.$$
Now we do the inner product operation with $\psi_m(t)$, then
$$\dot{a}_m(t)+\sum_na_n(t)e^{i(\theta_n(t)-\theta_m(t))}\Big(\langle\psi_m(t)|PC(t)|\dot{\psi}_n(t)\rangle+
\frac{1}{2}\langle\psi_m(t)|P\dot{C}(t)|\psi_n(t)\rangle\Big)=0.$$
For an explicit presentation, we have
$\dot{a}_m(t)=p_m(t)a_m(t)+q_m(t), $ where
$$p_m(t)=-i\textmd{Im}\langle\psi_m(t)|PC(t)|\dot{\psi}_m(t)\rangle,$$
\begin{eqnarray*}
q_m(t)&=&-\sum_{n\neq m}a_n(t)e^{i(\theta_n(t)-\theta_m(t))}\Big(\langle\psi_m(t)|PC(t)|\dot{\psi}_n(t)\rangle+
\frac{1}{2}\langle\psi_m(t)|P\dot{C}(t)|\psi_n(t)\rangle\Big)\\
&=&\sum_{n\neq m}a_n(t)e^{i(\theta_n(t)-\theta_m(t))}\Big(\langle\dot{\psi}_m(t)|PC(t)|{\psi}_n(t)\rangle+
\frac{1}{2}\langle\psi_m(t)|P\dot{C}(t)|\psi_n(t)\rangle\Big).
\end{eqnarray*}
Hence,
$a_m(t)=e^{\int_0^tp_m(s)ds}\Big(\int_0^tq_m(s)e^{-\int_0^sp_m(r)dr}ds+1\Big)$.
Because
$$|(\psi_m(t)|\psi(t))_t|=|a_m(t)|=\Big|\int_0^tq_m(s)e^{-\int_0^sp_m(r)dr}ds+1\Big|\ge
1-\Big|\int_0^tq_m(s)e^{-\int_0^sp_m(r)dr}ds\Big|,$$ and
\begin{eqnarray*}
\Big|\int_0^tq_m(s)e^{-\int_0^sp_m(r)dr}ds|&\le&
\int_0^t \sum_{n\neq m}|a_n(s)||\langle\dot{\psi}_m(s)|PC(s)|{\psi}_n(s)\rangle|ds\\
&&+\frac{1}{2}\int_0^t \sum_{n\neq m}|a_n(s)||\langle\psi_m(s)|P\dot{C}(s)|\psi_n(s)\rangle|ds\\
&\le& \int_0^t \left(\sum_{n\neq m}|a_n(s)|^2\right)^{1/2} \left(\sum_{n\neq m}|\langle\dot{\psi}_m(s)|PC(s)|{\psi}_n(s)\rangle|^2\right)^{1/2}ds\\
&&+\frac{1}{2}\int_0^t \left(\sum_{n\neq m}|a_n(s)|^2\right)^{1/2}\left(\sum_{n\neq m}|\langle\psi_m(s)|P\dot{C}(s)\psi_n(s)\rangle|^2\right)^{1/2}ds\\
&\le&\int_0^t\|\dot{\psi}_m(s)\|_{s}ds+\frac{1}{2}\int_0^t\|C(s)P(\dot{C}(s))^{\dag}P^{\dag}\psi_m(s)\|_{s}ds\\
&\le&\int_0^t\|\dot{\psi}_m(s)\|_{s}ds+\frac{1}{2}\int_0^t\|C(s)\dot{C}(s)\psi_m(s)\|_{s}ds\\
&\le&\int_0^t\|(PC(s))^{1/2}\|\left(\|\dot{\psi}_m(s)\|+\frac{1}{2}\|C(s)\dot{C}(s)\psi_m(s)\|\right)ds.
\end{eqnarray*}
Hence, when $V(T)<\varepsilon$, we have
$1-|(\psi_m(t)|\psi(t))_t|<\varepsilon$ for all $[0,T]$. $\Box$

\section{An example}

Put
$$P=\left(
\begin{array}{cc}
 0 & 1\\
1 &0
\end{array}
\right), T\left(
\begin{array}{c}
 x\\
y
\end{array}
\right)=\left(
\begin{array}{c}
 \overline{x}\\
\overline{y}
\end{array}
\right),
$$
where $\overline{a}$ means the conjugate of $a$. It is easy to see that the Hamiltonian
$$H(t)=\left(
\begin{array}{cc}
 s(t)e^{i\alpha(t)} & s(t)\\
s(t) & s(t)e^{-i\alpha(t)}
\end{array}
\right),
$$
is $PT$-symmetric, where $s(t)$ and $\alpha(t)$ are continuous real-valued
functions on $[0,\infty]$ such that $\cos\alpha(t)\geq \frac{1}{2}$ for all $t$ and $\dot{\alpha}(t)$ is continuous on $[0,\infty]$. It can be computed that $H(t)$ has  eigenstates $$\psi_1(t)=\frac{1}{\sqrt{2}}\left(\begin{array}{c}
 e^{-i\alpha(t)/2}\\
-e^{i\alpha(t)/2}
\end{array}
\right) \mbox{ and }\psi_2(t)=\frac{1}{\sqrt{2}}\left(
\begin{array}{c}
 e^{i\alpha(t)/2}\\
e^{-i\alpha(t)/2}
\end{array}
\right)$$ with corresponding eigenvalues $0$ and
$2s(t)\cos\alpha(t)$, respectively. Take
$$C(t)=\frac{1}{\cos\alpha(t)}\left(
\begin{array}{cc}
 i\sin\alpha(t) & 1\\
1 &-i\sin\alpha(t)
\end{array}
\right).$$
Then $\{C(t),P,T\}$ becomes a $C(t)PT$-frame for $H(t)$ such that $\{\psi_1(t),\psi_2(t)\}$ is an orthonormal basis for $\C^2$ with respect to $C(t)PT$-inner product $(\cdot|\cdot)_t$.

Now let we check the adiabatic approximation for $H(t)$. Let $0<\varepsilon<1$.
 If $\int_0^T
|\dot{\alpha}(s)|ds<\frac{\varepsilon}{6}$, we compute that
\begin{eqnarray*}
V(T)&=&\int_0^T\|(PC(s))^{1/2}\|\left(\|\dot{\psi}_m(s)\|+
\frac{1}{2}\|C(s)\dot{C}(s)\psi_m(s)\|\right)ds\\ &\leq&\int_0^T
4\left(\frac{|\dot{\alpha}(s)|}{2}+|\dot{\alpha}(s)|\right)ds\\
&<&\varepsilon.
\end{eqnarray*}
By Theorem 6, if  $\psi(t)$ is  a solution to (5)  with
$\psi(0)=\psi_m(0)$, then
$1-|(\psi_m(t)|\psi(t))_t|<\varepsilon$
all $[0,T]$.

\section*{ACKNOWLEDGMENTS}

This subject was supported by the National Natural Science Founds of China (no. 11171197), the Fundamental Research Founds for the Central Universities (GK201104010) and the Innovation Founds for Graduate Program of Shaaxi Normal University (2011CXB004).

\end{document}